\documentclass[11pt,a4paper]{article}  

\usepackage[latin1]{inputenc} 
\usepackage{geometry}          
\usepackage[english]{babel}    
\usepackage{epsfig}
\usepackage{setspace}

\title{\textbf{The Earth as an extrasolar planet:\\ The vegetation spectral signature today and during the last Quaternary climatic extrema}}
         
\author{Luc Arnold \\ \small Observatoire de Haute Provence CNRS INSU, 04870 Saint-Michel-l'Observatoire, France \\ \small (Luc.Arnold @ oamp.fr)\\
\and Fran\c cois-Marie Br\'eon \\  \small	CEA-DSM-LSCE, 91191 Gif-sur-Yvette, France \\
\and Simon Brewer \\  \small	CEREGE, BP 80, 13545 Aix-en-Provence Cedex 04, France \\
}
\date{Submitted 4th September 2008, accepted 31th December 2008, \\ International Journal of Astrobiology \\ http://journals.cambridge.org/action/displayJournal?jid=IJA }

\begin{document}
\doublespacing

\maketitle 

\begin{abstract}
The so-called Vegetation Red-Edge (VRE), a sharp increase in the reflectance around $700\ nm$, is a characteristic of vegetation spectra, and can therefore be used as a biomarker if it can be detected in an unresolved extrasolar Earth-like planet integrated reflectance spectrum.  Here we investigate the potential for detection of vegetation spectra during the last Quaternary climatic extrema, the Last Glacial Maximum (LGM) and the Holocene optimum, for which past climatic simulations have been made. By testing the VRE detectability during these extrema when Earth's climate and biomes maps were different from today, we are able to test the vegetation detectability on a terrestrial planet different from our modern Earth. Data from the Biome3.5 model have been associated to visible GOME spectra for each biome and cloud cover to derive Earth's integrated spectra for given Earth phases and observer positions. The VRE is then measured. Results show that the vegetation remains detectable during the last climatic extrema. Compared to current Earth, the Holocene optimum with a greener Sahara slightly increases the mean VRE on one hand, while on the other hand, the large ice cap over the northern Hemisphere during the LGM decreases vegetation detectability. We finally discuss the detectability of the VRE in the context of recently proposed space missions.

\end{abstract}


\section{Introduction}            
\begin{figure}
   \centering
   \includegraphics[width=10cm]{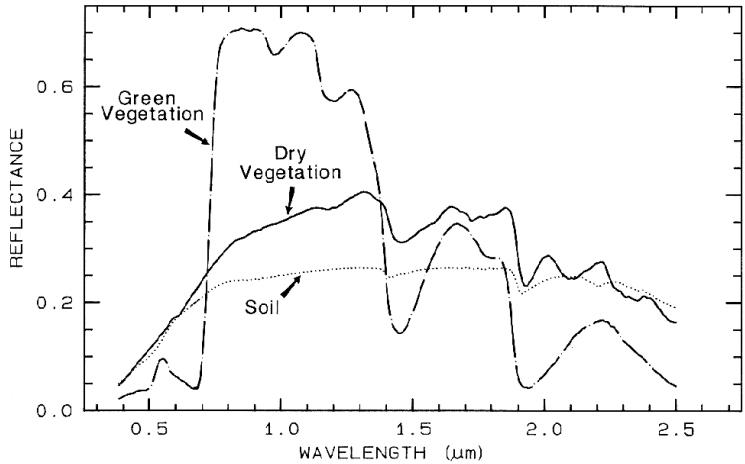}
   \caption{Reflectance spectra of photosynthetic (green) vegetation, non-photosynthetic (dry) vegetation and soil (from \cite{clark99}).
   The so-called vegetation red edge (VRE) is the green vegetation reflectance strong variation from $\approx5\%$ at $670\ nm$ to $\approx70\%$ at $800\ nm$.
   Note that the spectrum of a non-vegetated soil also shows a higher reflectance in the near-IR than in the visible.}
\label{vegetation}
\end{figure}

One of the most challenging goals of the future space missions is to detect life on extrasolar planets. These missions are designed to image a planet as a single dot floating beside its parent star, using high-contrast imaging techniques. Photons from the unresolved planet will feed a low-resolution spectrograph and give first insights into planet's chemistry. Recent works suggest that Earth's vegetation is detectable in the Earth integrated spectrum at visible wavelengths \cite{arnold2002,woolf2002,arnold2008}, assuming extraterrestrial plants, if they exist, are similar to Earth's plants. Our green vegetation indeed has a reflectance spectrum showing a sharp increase - the so-called vegetation red edge (VRE) - around $700\ nm$ (Fig. \ref{vegetation}) that could be considered as a possible global biomarker.

A question arising from these results is whether vegetation was also detectable when the (early) Earth looked significantly different than today. In other words, if an Earth-like planet with a climate different from the modern Earth's climate is detected one day, will we be able to detect vegetation on it? To answer this, we investigate the vegetation spectral detectability during the last Quaternary climatic extrema, for which past climate simulations have been made by general circulation models (GCMs). This simulated climate is used as a driver for an equilibrium vegetation model, providing biome maps of the Earth at these epochs.

The Last Glacial Maximum (LGM) occurred about 21 kyr (kilo-year) ago. Temperatures globally of the order of $4^\circ C$ colder than today \cite{braconnot2007} were responsible for a large extent of sea and continental ice sheets, especially in the northern Hemisphere. Sea level was also significantly lower ($-121\pm 5$ m \cite{fairbanks1989}) resulting in larger emerged lands.

During the Holocene optimum (6 kyr before present, hereafter kyrBP), Earth's northern Hemisphere was  $\approx0.5^\circ C$ warmer than today \cite{braconnot2007}. The sea level was rising, although still slightly lower than today, because the amount of water released from continental ice was partially compensated by an isostatic response of the continents, \textit{i.e.} a rise of the continents following the deglaciation of the northern Hemisphere. The Sahara was more vegetated than today \cite{ritchie1987,bonfils2001}, its desertification being the main large-scale change in land cover for the last 6000 years \cite{claussen1999}.

Here, based on a combination of biomes and simulated cloud cover maps for the periods mentioned above, associated with 300 to 800 nm spectra for each biome and cloud, we build Earth integrated spectra, for given Earth phases and observer positions. The vegetation spectral signature is then extracted from these spectra. The first part of the paper describes the model of the Earth from which we derive global reflectance spectra. The second part of the paper discusses the results. Note that some results showing the rotating Earth and corresponding instantaneous spectrum are available on-line as animated gif files (http://www.obs-hp.fr/$\sim$arnold/results/2009\_IJAstrobio/VRE.html).

%
%

\section{Main inputs for the Earth model}           
\subsection{Biomes maps from \textsl{Biome3.5} model for 0, 6 and 21 kyrBP} 

\begin{figure}
   \centering
   \includegraphics*[viewport=0 0 520 249]{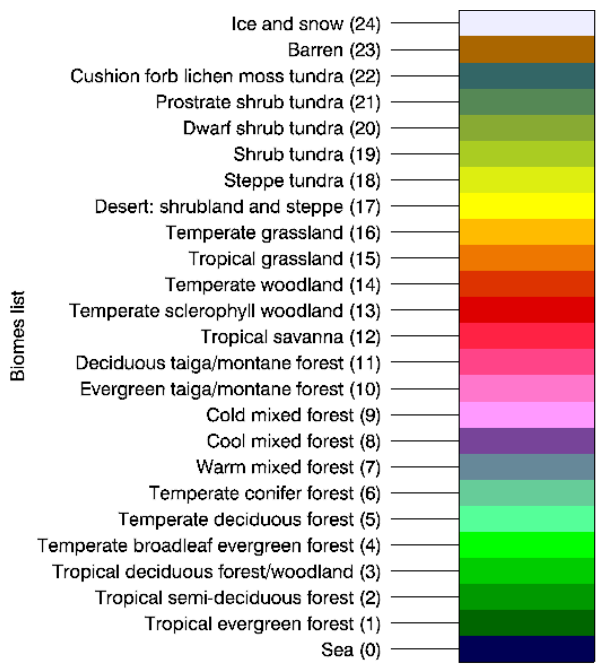}
   \caption{Map legend and biomes index.}
\label{biomes_legend}
\end{figure}

\begin{figure}
   \centering
   \vspace{-1,5cm}
   \fbox{\scalebox{0.75}{\includegraphics*[viewport=0 1 510 256]{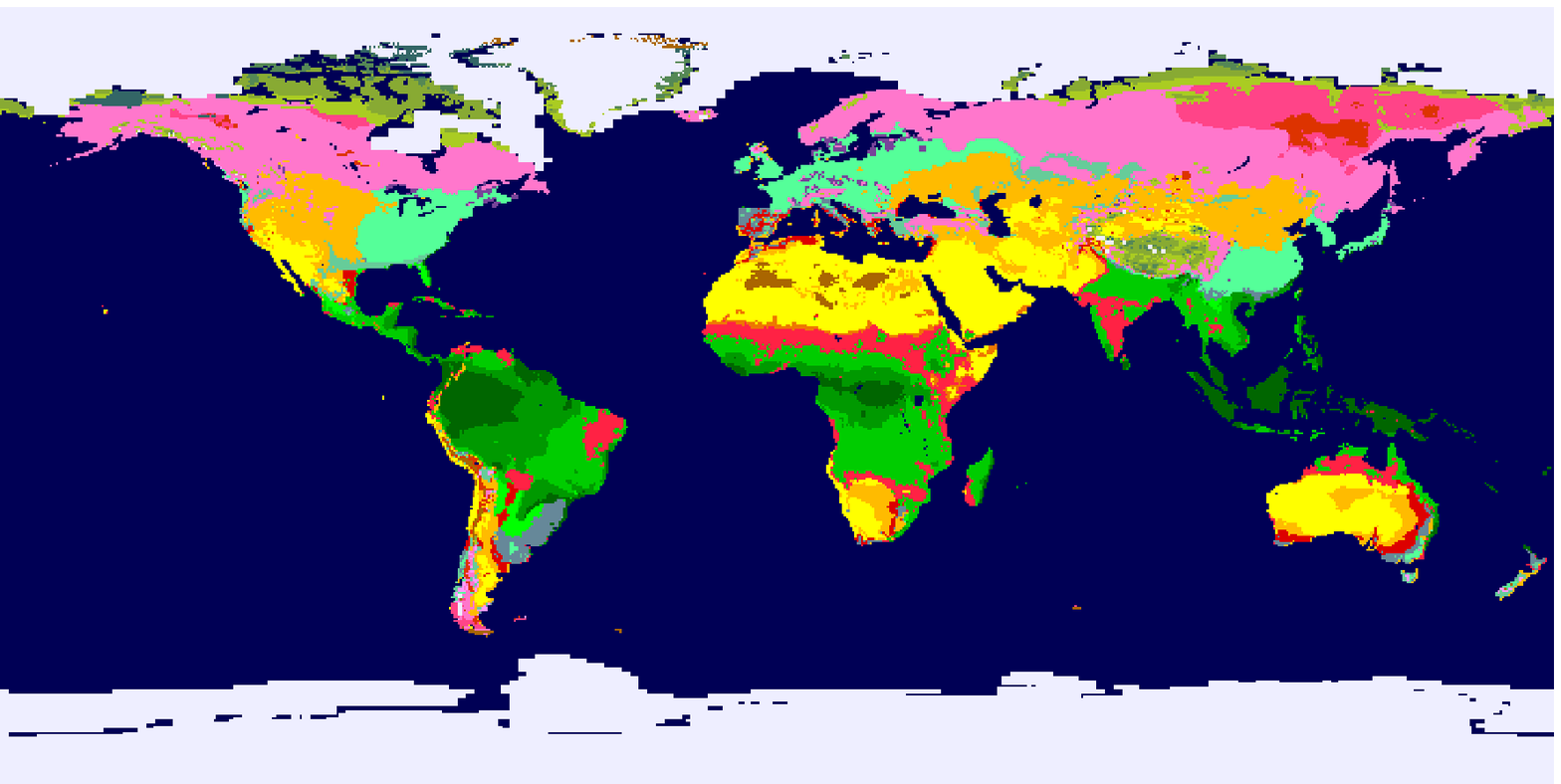}}}\\
   \fbox{\scalebox{0.75}{\includegraphics*[viewport=0 1 510 256]{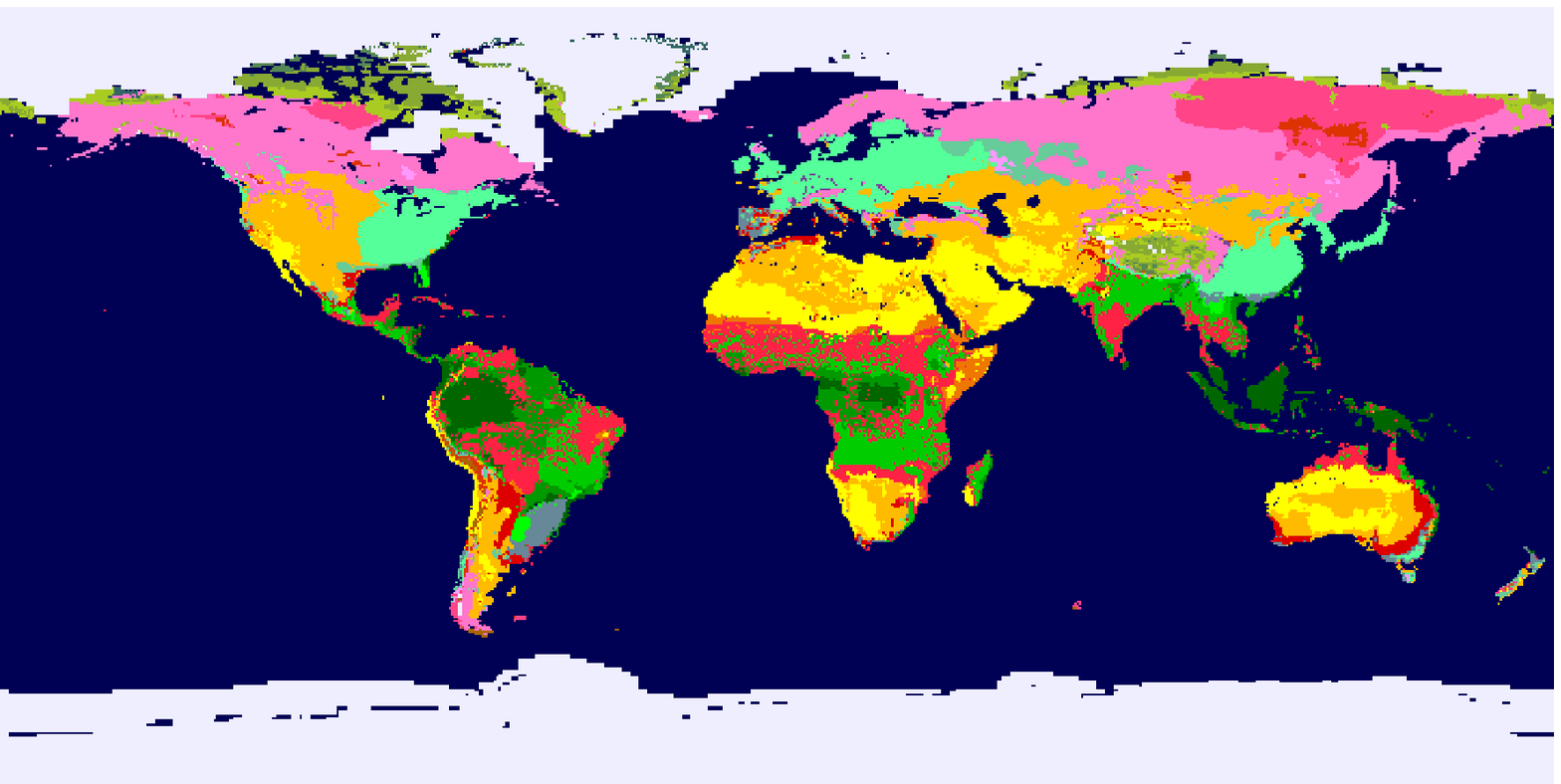}}}\\
   \fbox{\scalebox{0.75}{\includegraphics*[viewport=0 1 510 256]{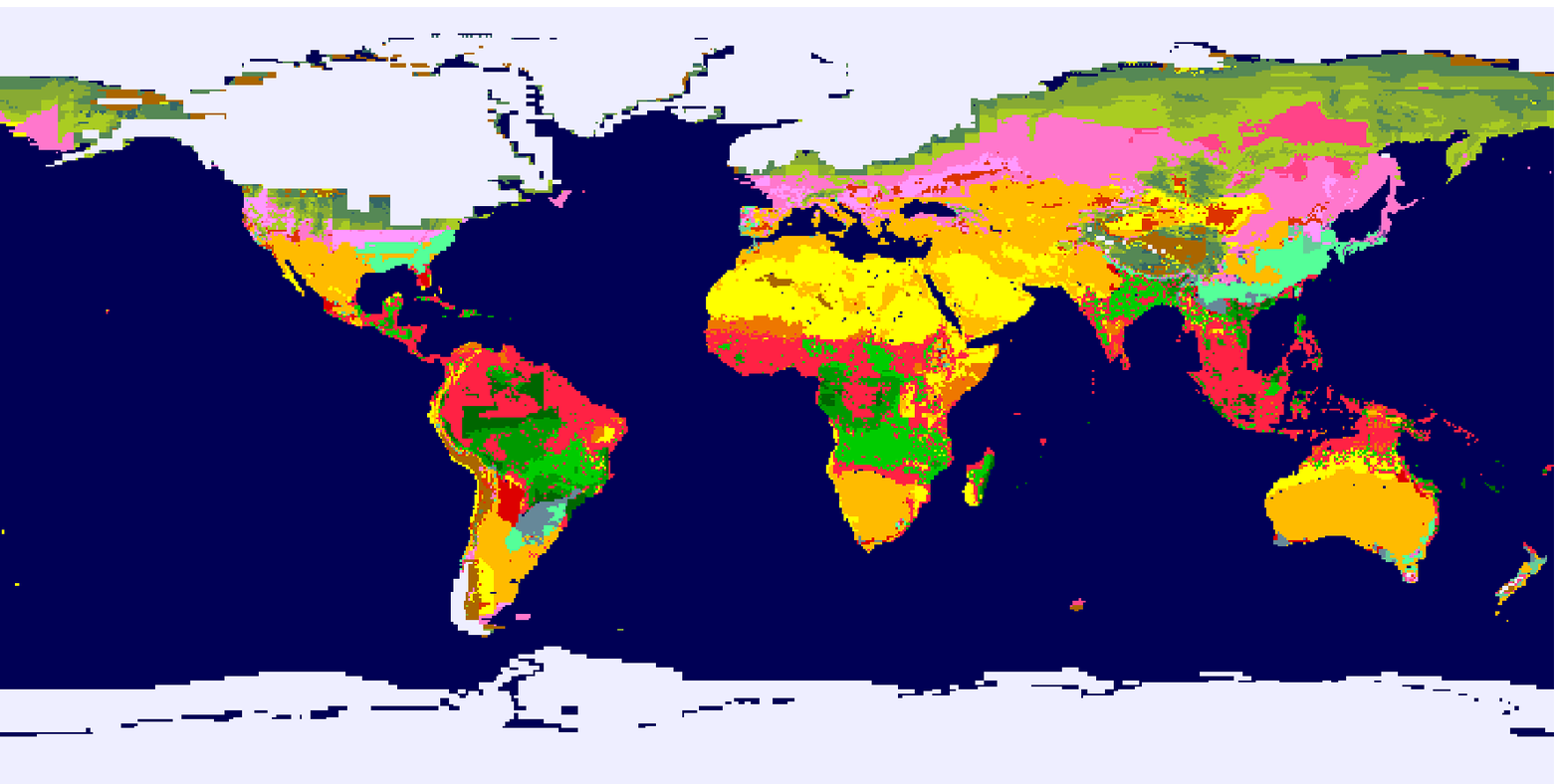}}}
   \caption{Maps of biomes today (top), 6 kyrBP during the Holocene optimum (middle) and 21 kyrBP  during the LGM (bottom), with continental and sea ices. Note that 6 kyrBP, a large part of the Sahara displayed a temperate grassland biome and almost no barren regions, while today and at 21 kyrBP, it consists of barren and desert (shrubland and steppe) landscapes. }
\label{biomes_maps}
\end{figure}

The equilibrium terrestrial biosphere model BIOME3 \cite{haxeltine1996} is a process based vegetation model simulating biogeography and biogeochemistry. Model inputs consist of monthly climate variables (temperature, precipitation, and sunshine) data, minimum annual temperature and atmospheric CO2 concentration. A coupled hydrological model calculates moisture availability on the basis of a soil texture class. The model simulates the productivity and cover (Leaf Area Index) of a set of plant functional types. The comparative dominance of these types is then used to estimate the dominant biome for a given grid cell. A total of 24 biomes may be simulated, including barren ground and snow/ice cover (Fig. \ref{biomes_legend} and \ref{biomes_maps}). 

The model was run for the three periods of interest: modern, Holocene optimum and LGM Earth (Fig. \ref{biomes_legend} and \ref{biomes_maps}). For the modern period, gridded climate data on a 0.5� grid was taken from \cite{leemans91}, and atmospheric CO2 concentration was set to a pre-industrial level of 280 ppm. For the two past periods, climate data was obtained from simulations run for under the Paleoclimate Modelling Intercomparison Project\footnote{Paleoclimate Modelling Intercomparison Project (PMIP), http://pmip.lsce.ipsl.fr/.}. We used data from the UK Meteorological Office General Circulation Model (UKMO GCM), run under climatic forcings for these two periods \cite{hewitt96, hewitt97}. The GCM output was downscaled to the same resolution grid as the modern period by calculating anomalies for each past period. The GCM control run value was subtracted from the past value for each GCM grid point, and then these anomalies were added to the modern climatology grid points. Atmospheric CO2 concentration was set to 265 ppm and 180 ppm for the Holocene optimum and LGM respectively. 

The results obtained agree well with observational evidence of past vegetation change \cite{prentice2000}. The mid-Holocene is dominated by a greening of the Sahara, and a poleward shift of the high latitude treelines. The LGM is characterised by a large expansion of steppe and grassland vegetation at mid-latitudes, with a near disappearance of temperate forest. There is also the fragmentation of tropical forest and widespread expansion of tundra vegetation at high latitudes. There are, however, some disagreements with the data, for example the expansion of grassland in Australia.

\subsection{Biome spectra} 
\label{sect_biome_spectra}
\begin{figure}
   \centering
   \vspace{-1,5cm}
   \includegraphics[width=12.5cm]{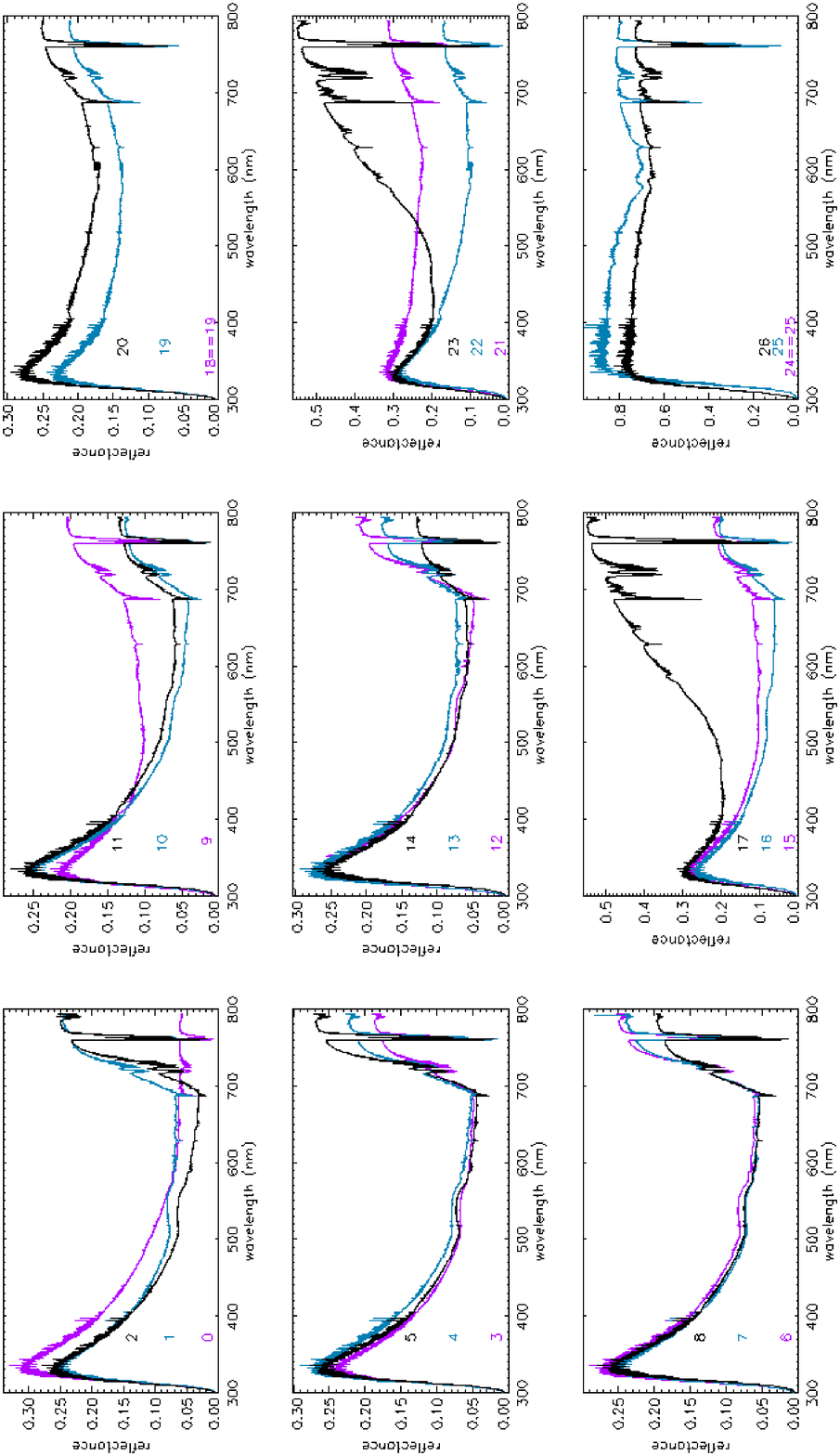}
   \caption{Biomes spectra measured by GOME the 18, 19 and 20th of May 1999. The spectra have been recorded at air-mass (here total path length through the atmosphere) ranging from 2 to 2.6 depending
   on solar and spacecraft line-of-sight angles. See Fig. \ref{biomes_legend} for the biome legend. Note that all plots do not have the same vertical scale. }
\label{GOME_spectra}
\end{figure}

We used a collection of GOME spectra taken the 18, 19 and 20th of May 1999, corresponding to springtime and the middle of the greenup period in the northern Hemisphere. In May the average NDVI (Normalized Differential Vegetation Index) is just half of its maximum value reached in July\footnote{http://earthobservatory.nasa.gov/Newsroom/NewImages/Images/PIA04335.gif}. 
Sun declination is $\approx+19.7^\circ$ for the observing dates in May. It is fall in the southern Hemisphere and the polar night is present at latitudes below $-70.3^\circ$.
The GOME spectra are reflectances obtained from the ratio of a given ground pixel radiance $R$ over solar irradiance $I_{\odot}$, 
\begin{equation}
  r_{b}=\frac{\pi\ R}{ I_{\odot} \cos\theta_s}
  \label{r_b}
\end{equation}
where $\theta_s$ is the solar zenit angle.
A reference spectra for each biome $r_b$ is extracted from the GOME data (Fig. \ref{GOME_spectra}). In order to find a given biome spectrum with minimal cloud contamination, the algorithm choses the spectrum with the lowest signal in the interval 450 to 500 nm, considering that the signal in this spectral range rapidly increases with the presence of clouds. This algorithm fails for three types of surfaces: ocean, snow/ice and cloud. These spectra are chosen 'by hand'. 

It must also be noted that this algorithm chooses the most vegetated pixel for a given biome: if a cloudless pixel is a mixture of sand and vegetation, it will be ignored in favor of a pixel with more vegetation, which will display a lowest signal in the spectral interval mentioned. This suggests that the reference pixel (and spectrum) for each biome is probably biased toward the most vegetated pixel of a given biome area. 

Lastly, since it is spring in the northern Hemisphere and fall in the southern Hemisphere, the reference spectrum for the biomes of deciduous species will be preferentially picked up in the 'greener' northern Hemisphere. Therefore, our simulations will be more relevant for reconstruction of the northern Hemisphere rather of the southern Hemisphere, with a probable overestimate of the vegetation signal if the southern Hemisphere is mainly in view.

\subsection{Cloud cover}

\subsubsection{Modern cloud cover reflectance from ISCCP data} 
To build a realistic Earth's spectrum, we need both the cloud cover map and the reflectance map of that cloud cover, $r_{cc}$. We used cloud covers from the International Satellite Cloud Climatology Project (ISCCP\footnote{http://isccp.giss.nasa.gov/}). We selected only data for the month of May from the 1999 to 2004 Vis-IR observations, to be consistent with the GOME spectra. 
Since low, middle and high altitude clouds have different typical optical depth and reflectances, we build the cloud cover reflectance $r_{cc}$ from the sum of the three different altitudes cloud cover maps $cc_L$, $cc_M$ and $cc_H$, weighted by their albedo, $a_L$, $a_M$ and $a_H$ respectively 0.69, 0.48 and 0.21 \cite{manabe64}, and normalized to the albedo of low altitude clouds $a_L$, i.e.

\begin{equation}
 r_{cc}=cc_L + \frac{a_M}{a_L} cc_M + \frac{a_H}{a_L} cc_H.
 \label{r_cc}
\end{equation}

\begin{figure}
   \centering
   \fbox{\scalebox{1}{\includegraphics*[viewport=0 1 510 490,width=14cm,height=7.5cm]{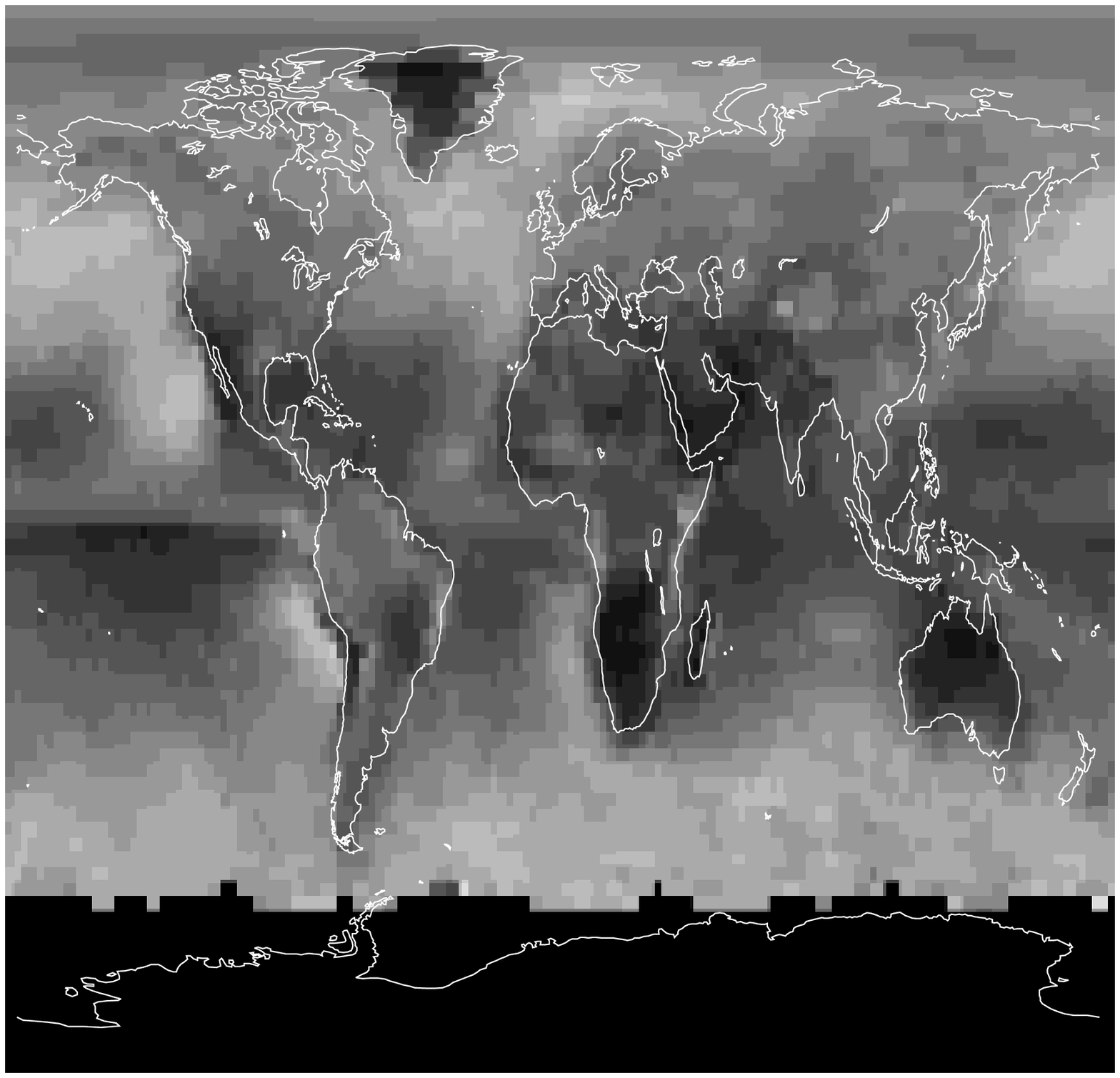}}}
   \includegraphics*[angle=-90,width=15cm]{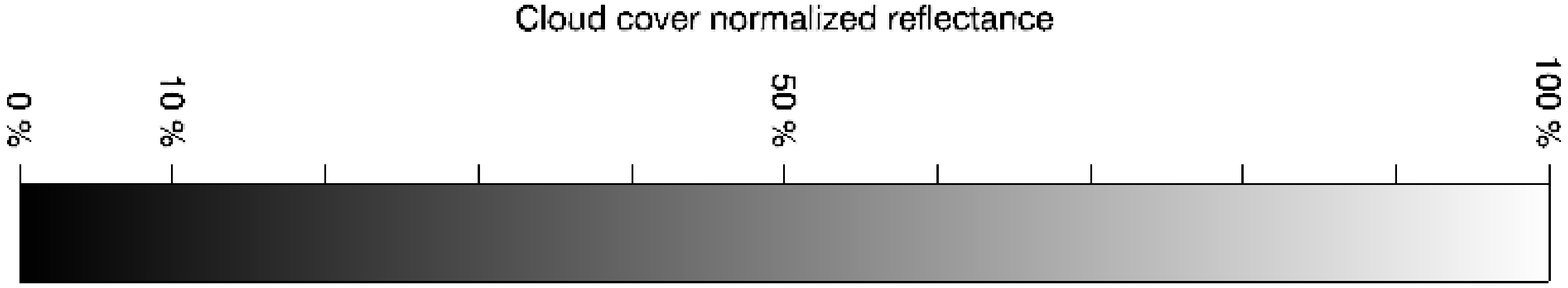}
   \caption{Normalized reflectance map of the cloud cover estimated from ISCCP data on low, mid and high cloud cover maps and clouds albedo \cite{manabe64} for the mean of cloud cover for month of May for six years from 1999 to 2004. No data are available for the Antarctica because this region is always in the night in May. Quantitative data are given in Table \ref{data_cc}.
}
\label{cc_ref_isccp}
\end{figure}

The result from this formula is that the three different types of clouds contribute to a given pixel's reflectance proportionally to their cover over that pixel. The formula also can lead to a few pixels with reflectance above 1, which are then simply clipped to 1. The obtained map (Fig. \ref{cc_ref_isccp}) is normalized to the albedo of low clouds for which we have a reference spectrum from the GOME data. 

\begin{table*}
\caption{Quantitative data about cloud cover normalized reflectance from real data (ISCCP) and model (biome3.5).}             
\label{data_cc}      
\centering          
\begin{tabular}{llccc}     
\hline\hline       
  				&  Modern Earth  &   Modern Earth & Holocene optimum 	& LGM \\
  				& (ISCCP)        &    (biome3.5)  & (biome3.5, 6 kyrBP) &(biome3.5, 21 kyrBP)\\ 
\hline                    
mean $R_{cc}$  			& 0.46  &		0.40	 	& 0.40		& 0.39 \\
max  $R_{cc}$  			& 0.86  &		1.00	 	& 1.00		& 1.00 \\
min  $R_{cc}$  			& 0.00  &		0.03     	& 0.03		& 0.05 \\
+60\char23 +90\char23 		& 0.51  &		0.50     	& 0.50		& 0.41 \\
+40\char23 +60\char23 		& 0.53  &		0.52     	& 0.52		& 0.58 \\
-40\char23 +40\char23 		& 0.38  &		0.22     	& 0.22		& 0.24 \\
-60\char23 -40\char23 		& 0.62  &		0.81     	& 0.81		& 0.75 \\
\hline                 
\end{tabular}
\end{table*}

\begin{figure}
   \centering
   \fbox{\scalebox{1}{\includegraphics*[viewport=0 1 510 490,width=14cm,height=7.5cm]{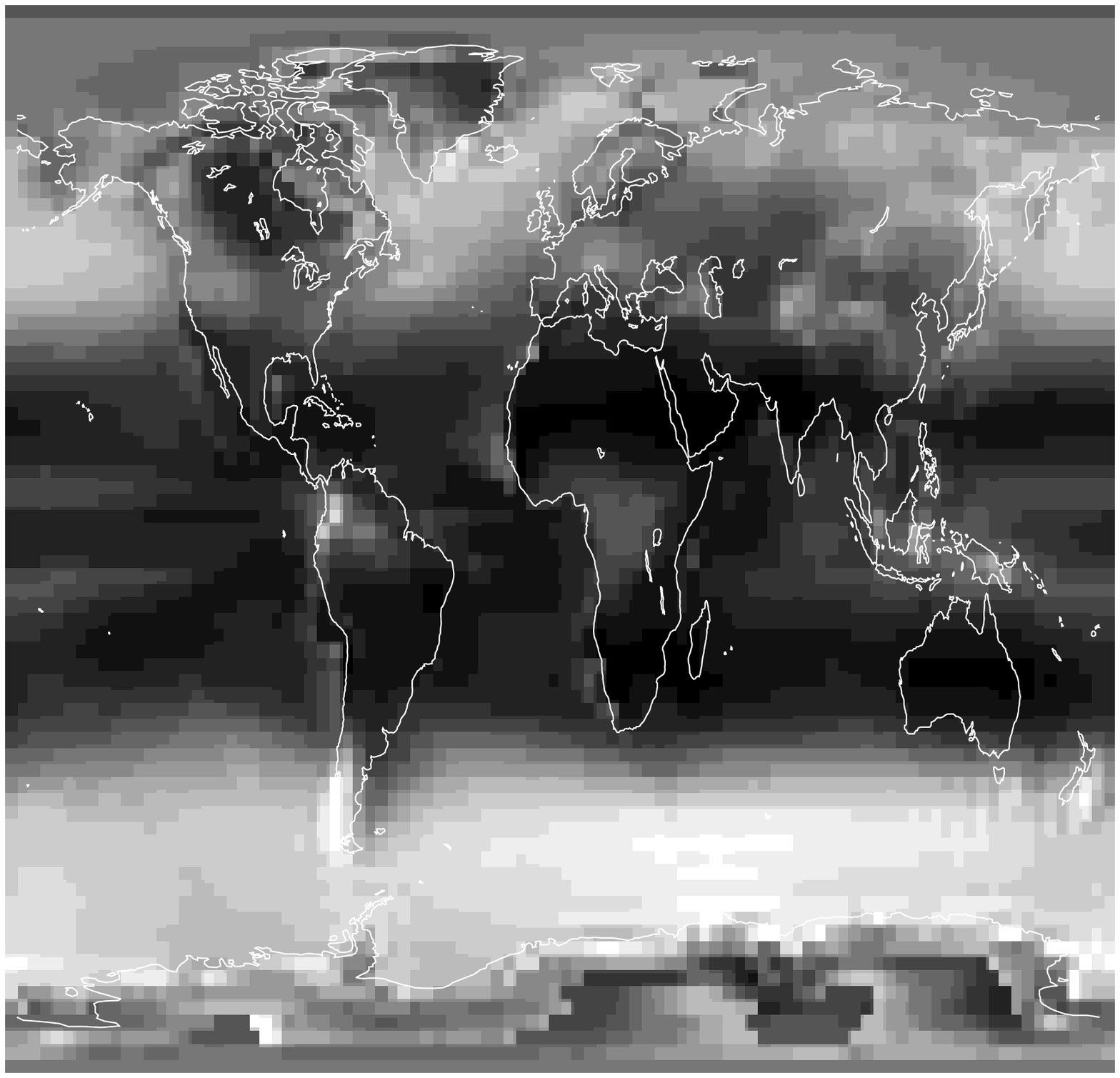}}}\\
   \fbox{\scalebox{1}{\includegraphics*[viewport=0 1 510 490,width=14cm,height=7.5cm]{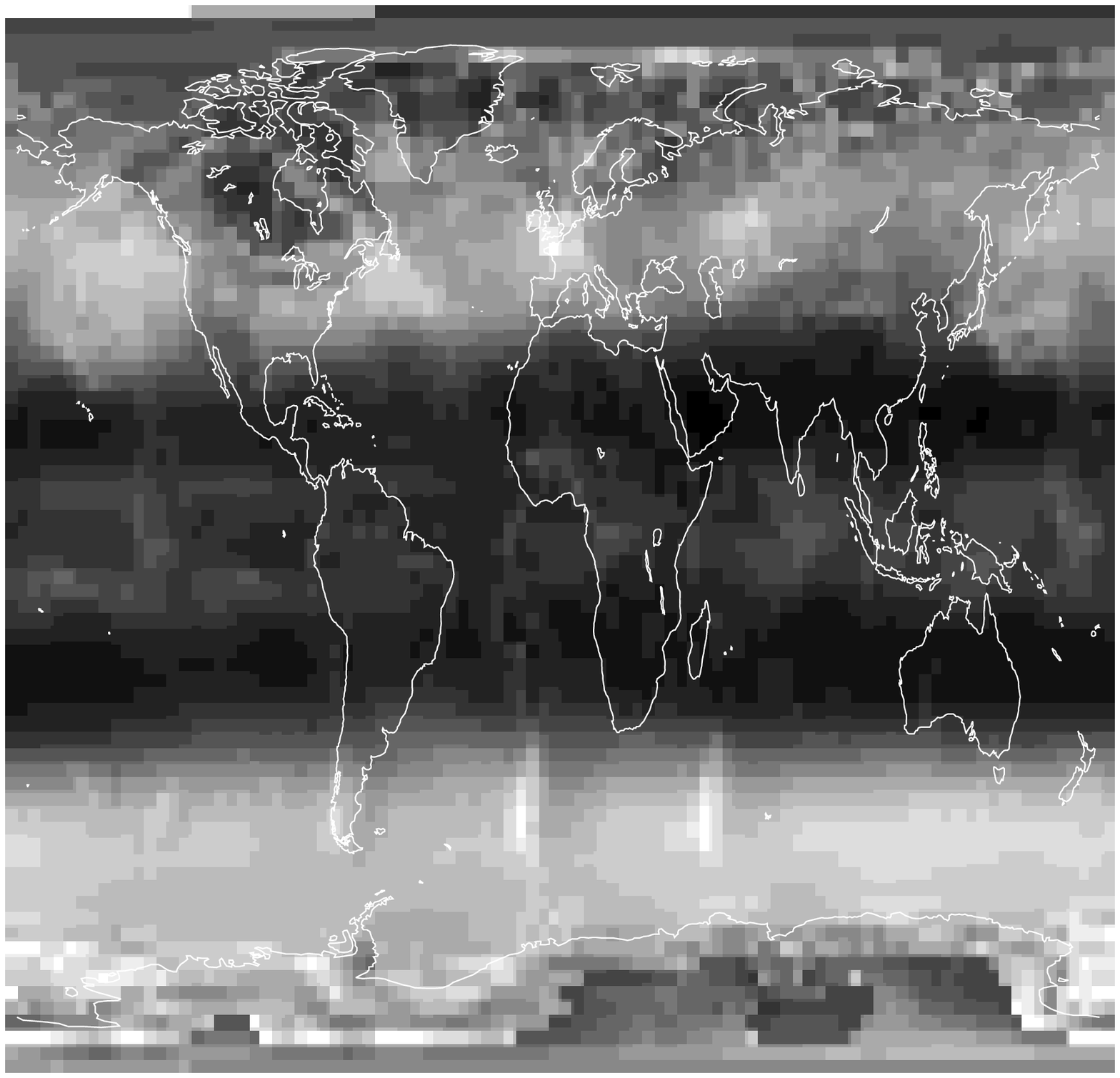}}}\\
   \caption{Normalized reflectance maps of the cloud cover estimated from cloud cover simulated by the UKMO GCM, with the same grey-scale as for Fig.~\ref{cc_ref_isccp}, for modern Earth or at 6 kyrBP (top) and 21 kyrBP (bottom). 
   In the 21 kyrBP map, artifacts induced by the existence of a wavetrain resulting from the passage of the lower atmospheric circulation over the topography of South America are visible in the southern Hemisphere: The cloud cover is slightly overestimated in the middle of the Atlantic Ocean and at the South of the Madagascar island. Quantitative data are given in Table \ref{data_cc}.}
\label{cc_ref_biome35}
\end{figure}

\subsubsection{Cloud covers from \textsl{Biome3.5} model for 0, 6 and 21 kyrBP} 
Estimates of past changes in cloud cover were taken from the UKMO GCM, as used as input for the Biome 3.5 model. The 0 and 6 kyrBP cloud covers are the same (Fig.~\ref{cc_ref_biome35}). The simulated cloud cover at 21 kyrBP shows locally overestimated
cloud cover in the southern Hemisphere, caused by the existence of a wavetrain resulting from the passage of the lower atmospheric
circulation over the topography of South America (Fig.~\ref{cc_ref_biome35}). The simulated cloud cover output consists of 17 layers of clouds at given pressures from 1000 (ground) to 10 mbar (highest clouds). The model shows that the highest layers, at 30, 20 and 10 mbar, are very high clouds above Antarctica which are therefore not relevant for our study. The remaining 14 layers are quadratically summed into three layers of low, middle and high clouds, according the ISCCP classification versus cloud pressure\footnote{http://climate.msrc.sunysb.edu/cpt/isccp-cltypes.gif} (i.e. low clouds for pressures from 1000 to 680 mbar, middle clouds from 680 to 440 mbar, and high clouds above 440 mbar). A simple sum of the 14 layers into 3 would mean that the clouds in each layer contribute linearly and cumulatively to the total cloud cover over a given pixel. This is not realistic and leads to overestimation of cloud cover (many pixels have a cloud cover significantly larger than 100\%). Using a quadratic sum allows the superimposition of layers to be taken into account, leading to much more realistic results when the 0 kyrBP model is compared with real ISCCP data. The cloud reflectance maps are calculated following Eq. \ref{r_cc} as done with the ISCCP data.

\section{Disk-averaged spectra from the model and extraction of the vegetation signal}
\label{method}
A 2-layer map of the Earth is created with biome and cloud cover for each pixel. For ocean, 2-layers are used too, but the biome is replaced with the sea ice cover. 
Once this composite map is built, we proceed with an orthographic projection where the central longitude and latitude corresponds to the position chosen for the virtual distant observer. A realistic image of the Earth -i.e. a radiance map-, with limb darkening and phase, is finally obtained by multiplying the previous projection by a mask of a white lambertian sphere at the desired phase. We consider indeed a lambertian Earth, i.e. all biomes, ocean and clouds have an isotropic reflectance. Thus no specular reflection on ocean surface is taken into account. Specular reflection has indeed an albedo of 0.02, moreover attenuated by lands and clouds \cite{breon06}. We estimate from \cite{breon06} that specular reflection would increase the overall Earth albedo by $\approx0.01$ and therefore has a negligible quantitative impact.

We assume that a lambertian approximation of biomes reflection, with respect to real anisotropic BRDF, has negligeable impact on the global spectrum. Specular reflection of the Sun in ocean would make the Earth brighter and consequently decrease the relative signal of vegetation. Our model also considers  all clouds as optically thick, according to ISCCP mean optical thickness of clouds\footnote{ISCCP data analysis http://isccp.giss.nasa.gov/climanal1.html} equal to 3.86.
We understand that to derive more accurate spectra, our model would probably benefit from radiative transfer calculations, an implementation of the sun glint effect and the BRDFs of all scattering surfaces included in the model (biomes, oceans, clouds).

The global spectrum is computed from the projection of the Earth described just above: The global Earth Reflectance spectrum $ER(\lambda)$ is a linear combination of the spectra of each biome, ocean (ice-free or partly covered by sea ice) and clouds, weighted by their respective number of pixels and integrated radiance in the projected image, a pixel being itself a combination of two or three elements (biome and cloud, or sea, sea ice and cloud). The result is normalized to the flux integrated in the image of a white lambertian sphere observed with the same phase, giving the result spectrum in terms of apparent albedo $A_v$ \cite{qiu2003}.

We want to measure the height of the VRE embedded in Earth reflectance $ER(\lambda)$, considered as an example of a large-scale ground-based signature of life. We want to remove all \textit{a priori} known biases, these biases being due to atmospheric molecules of biotic or abiotic origin, or due known minerals spread over large land surfaces. The VRE is quantitatively approximated by the ratio \cite{arnold2002, hamdani2006}
\begin{equation}
VRE=\frac{r_I-r_R}{r_R} \label{VRE}
\end{equation}
where $r_I$ and $r_R$ are the near-infrared (NIR) and red reflectances integrated over spectral domains defined on both sides of the Red Edge. This formula gives correct results only if the spectrum has been cleaned
from known sources of atmospheric bias, as described above. There are at least three sources of bias that can be corrected a priori and that have been considered here. 

The first is the absorption of atmosphere molecules. Ozone has a weak but wide absorption over the full visible range (Chappuis band), and absorption is non-zero at the wavelengths of interest for the VRE. The Earth spectrum is thus divided by a reference ozone absorption spectrum, extracted from a GOME spectrum of white clouds and adjusted to the data. Water vapor and oxygen absorptions are also corrected with MODTRAN\footnote{MODTRAN is distributed by Ontar Corporation.} reference spectra: the band depth of the reference spectra are optimized to minimize the residual of the fit, after which the spectrum is divided by the fit. These two molecules have lower impact on the VRE extraction than ozone because their absorption bands are relatively narrow, moreover the spectral domains over which the VRE is measured can be chosen between $H_2O$ and $O_2$ absorption bands.    

Scattering in the atmosphere is the second source of bias we take into account. Rayleigh and aerosol scattering produces a negative slope in the spectrum in the blue part of the spectrum. We consider here that the aerosol and the Rayleigh contribution to scattering are equal at $\lambda_e=700$ nm \cite{lena96}.

The third source of bias we take into account is the spectral signature of deserts that adds a smooth positive slope across the visible and near-IR domains where the VRE is measured \cite{arnold2003,arnold2008}, and the function defined to fit Rayleigh, aerosol and desert biases from 350 to 685 nm is thus written
\begin{equation}
f(\lambda)=\left(A_0 + {A_1\over \lambda^4+\lambda_e^{2.5}\lambda^{1.5}} + A_2\times s_d\right) \times s_{O_3}^{\ A_3}.  \label{fit}
\end{equation}
The spectrum $s_d$ is our model of desert reflectance, extracted from a GOME spectrum taken over Sahara and corrected for Rayleigh and ozone biases, and $s_{O_3}$ is the spectrum of ozone extracted from a GOME spectrum of white clouds. The $ER(\lambda)$ spectrum is then normalized to the vegetationless model $f(\lambda)$ to measure the VRE using the spectral domains [650:685nn] and [750:758nm].
It can be noted here that the correction of $H_2O$ and $O_2$ is only relevant for the fit function between 580 and 680 nm.

To test the robustness of these bias corrections, we built an Earth model where all pixels of continents were desert, with oceans, clouds, sea and continental ices remaining unchanged. We obtained a mean VRE of $0.1\%$ with a standard deviation of $0.2\%$ after an integration over a full Earth rotation, for an observer seeing mostly northern Hemisphere of an half-phase Earth (inclinasion $i=60\char23$, orbital phase $\phi=180\char23$ - eastern elongation). For identical geometrical parameters, the vegetated Earth model gives a VRE of $5.9\%$ (standard deviation of $2.9\%$) suggesting that the $f(\lambda)$ function well fits a vegetationless spectrum, i.e. meaning that the biases mentioned above are well corrected to allow the detection and the measurement of VRE values as low as $\approx1\%$.

\section{Results and discussion}

\begin{table*}
\caption{24h-averaged disk-averaged VRE (\%) and apparent albedo $A_v$ measured from our Earth model during springtime for different Earth phases and observer's positions. The angle $\phi$ is the orbital phase: $\phi=0\char23$ is
for Western maximum elongation, $\phi=90\char23$ for superior conjunction, $\phi=180\char23$ for Eastern elongation and $\phi=270\char23$ for inferior conjunction.}             
\label{VRE_albedo}      
\centering          
\begin{tabular}{llccc}     
\hline\hline       
(epoch $\longrightarrow$)	    	&  Modern  &   Modern  & Holocene optimum 	& LGM \\
																&  0 kyrBP  &  0 kyrBP   & 6 kyrBP          &21 kyrBP\\ 
(data $\longrightarrow$)				& ISCCP  &    Biome3.5  & Biome3.5          &Biome3.5\\ 
\hline                    
$i=45\char23, \phi=180\char23$&&&&\\
(eastern elongation, half-Earth)   &&&&\\
$<VRE>$  					   & 4.4     &            5.1 			& 5.8  								& 3.4 \\
$A_v $  					   & 0.368   &          0.323 		  & 0.320  					    & 0.359 \\
\hline
$i=45\char23, \phi=90\char23$ &&&&\\
(superior conjunction, $\approx$ full Earth)   &&&&\\
$<VRE>$  					   & 4.7   &            5.7 			& 6.7  								& 4.3 \\
$A_v $  					   & 0.347   &          0.293 			& 0.290  						& 0.328 \\
\hline
$i=45\char23, \phi=270\char23$ &&&&\\
(inferior conjunction, crescent Earth)   &&&&\\
$<VRE>$  					   & 2.5   &            2.4 			& 2.6  								& 0.8 \\
$A_v $  					   & 0.466   &          0.456 		& 0.455  					  	& 0.489 \\
\hline
$i=0\char23, \phi=180\char23$ &&&&\\
(Ecliptic North pole view, half-Earth) &&&&\\
$<VRE>$  					   & 4.3   &            4.3 			& 4.7  								& 2.0 \\
$A_v $  					   & 0.395   &          0.370 		& 0.368  							& 0.418 \\
\hline         
$i=90\char23, \phi=180\char23$&&&&\\
(eastern elongation, half-Earth)   &&&&\\
$<VRE>$  					   & 5.0     &            7.4 			& 8.4  								& 6.9 \\
$A_v $  					   & 0.312   &          0.237 		  & 0.234  					    & 0.255 \\
\hline
$i=135\char23, \phi=180\char23$&&&&\\
(eastern elongation, half-Earth)   &&&&\\
$<VRE>$  					   & 4.2     &            6.7 			& 7.0  								& 7.7 \\
$A_v $  					   & 0.311   &          0.233 		  & 0.232  					    & 0.231 \\
\hline
$i=180\char23, \phi=180\char23$&&&&\\
(Ecliptic South pole view, half-Earth)   &&&&\\
$<VRE>$  					   & 3.9     &            5.8 			& 5.8  								& 6.7 \\
$A_v $  					   & 0.324   &          0.255 		  & 0.255  					    & 0.249 \\
\hline

\end{tabular}
\end{table*}

The reconstructed scenes of the Earth, from which the results below are extracted, have been made using a Sun declination of $\delta=+19.7^\circ$, corresponding to the Earth-Sun configuration for May. The Earth phase viewed by the observer is reconstructed for several inclination $i$. The angle $\phi$ is the orbital phase of the Earth viewed from the location of the observer: $\phi=0\char23$ is for Western maximum elongation, $\phi=90\char23$ for superior conjunction, $\phi=180\char23$ for Eastern elongation and $\phi=270\char23$ for inferior conjunction.

In Table \ref{VRE_albedo}, the VRE and albedo values obtained from the model are given for the different periods. For the modern Earth, the albedo values obtained with the ISCCP data 
are higher than with Biome3.5, due to a lower cloud cover with Biome3.5, leading to a lower VRE. This suggests that the Biome3.5 model probably overestimates the VRE. Nevertheless, relative comparisons between Biome3.5 results remain consistent although the absolute values may be overestimated.

\begin{figure}
   \centering
   \includegraphics[width=13cm]{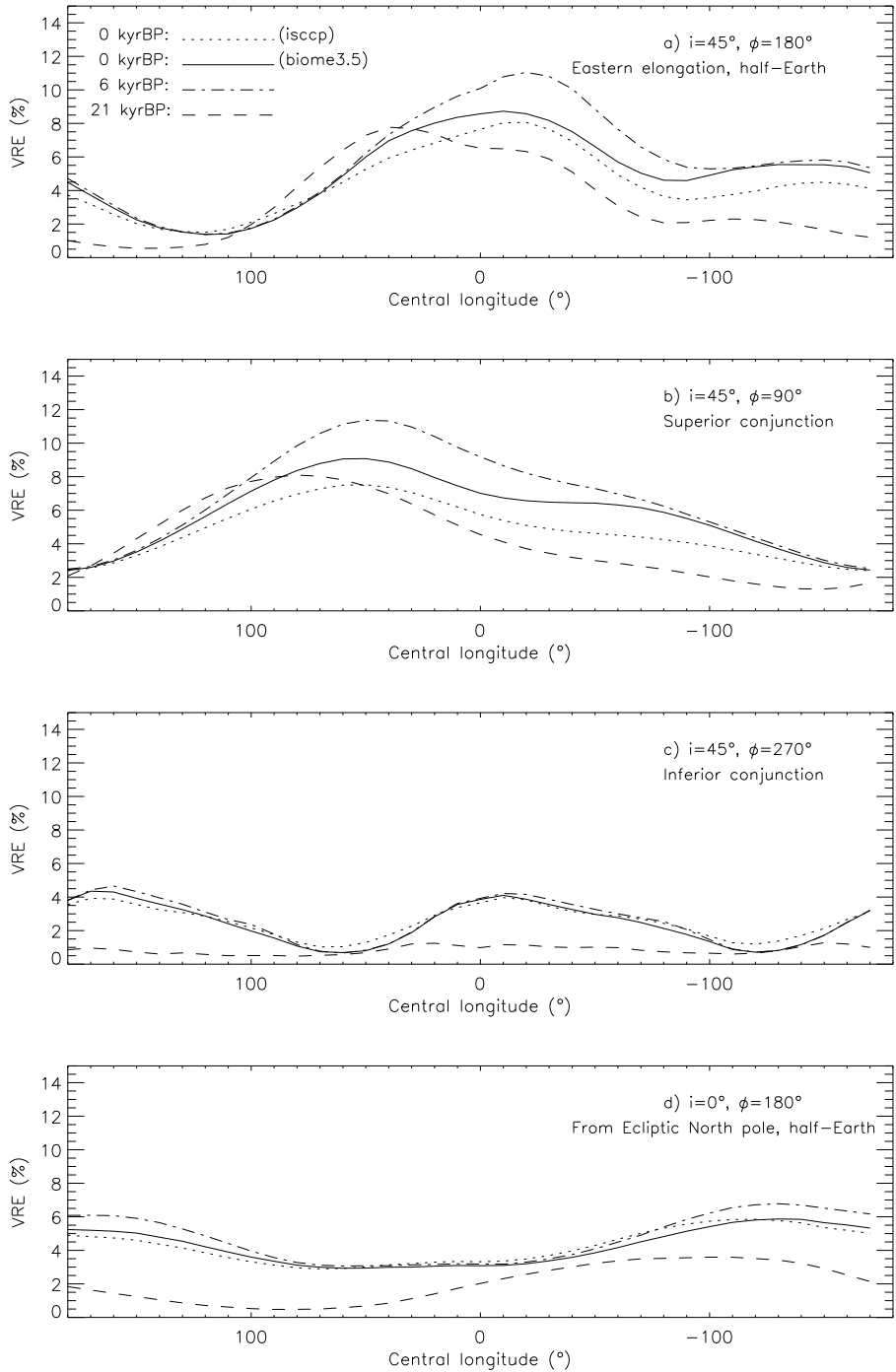}
   \caption{Vegetation Red Edge measured with Biome3.5 data for clouds, ice and biomes, and GOME spectra (and ISCCP for clouds at 0 kyrBP only). The VRE calculated with the ISCCP clouds and the BIOME3.5 model are in correct agreement, although the latter is typically $1\%$ higher. Central longitude is decreasing, according to the eastward Earth rotation. }
\label{VRE-24h-variation_nord}
\end{figure}

The main result from this study is that the climate differences between the Quaternary extrema and the modern Earth's climate have little impacts on the vegetation signal, since the spectral signature of vegetation is not washed out during the LGM and does not increase very much during the Holocene optimum. There
are nevertheless interesting differences in 'vegetation visibility' between these two extrema and the situation today, depending of the observer's position: When the northern Hemisphere is mainly in view, vegetation is indeed more visible during the Holocene optimum as shown in Fig.~\ref{VRE-24h-variation_nord} and Table \ref{VRE_albedo}: The VRE was typically 6\%, thus higher than today (5\%) or during the LGM (4\%). This is mainly due to a greener Sahara as shown in Fig. \ref{VRE-24h-variation_nord}(a). This is true for an observer with mostly the northern Hemisphere in view, i.e. inclination $i=45\char23$. In inferior conjunction configuration, the crescent-shaped Earth mainly shows its northern part. There are no significant differences between modern and Holocene Earths (VRE $\approx2\%$) and vegetation becomes almost undetectable for the Earth with the LGM large polar ice cap ($<1\%$). If the observer is above the northern pole (observing from ecliptic North pole) with in view a half-illuminated Earth, there are again no significant differences between modern and Holocene Earths (VRE $\approx4\%$). However, as the North pole ice is much more extended during the LGM, the signal of the vegetation is significantly weaker (only 2\%), due to the higher reflectance of a larger polar zone and the consequently smaller vegetated areas is view. 

The three bottom lines from Table \ref{VRE_albedo} give results for Earth views in which southern Hemisphere contribution is high, with inclination $i$ ranging from 90 to $180\char23$. As pointed out in Section \ref{sect_biome_spectra}, our model probably overestimates the VRE values of scenes with the southern Hemisphere mostly in view as our calibration was made during fall time (May) in that hemisphere and deciduous species should have a lower VRE than with the spectra used in the simulation that have been collected in the northern Hemisphere. Nevertheless, there are two trends to be mentioned: the first one is that the difference between Holocene and modern period becomes negligible, because the Sahara contribution decreases or even vanishes from these southern view points. The second remark is that, paradoxically, the southern Hemisphere VRE during the LGM is higher: this is induced by a greener Australia during the LGM, as we obtained a simulated wider distribution of temperate grassland and less desert in Australia using the Biome3.5 model (Fig. \ref{biomes_maps}) at the Holocene optimum. As discussed above, this does not agree with observations of past vegetation and may arise from an incorrectly simulated climate over this region. 

\begin{figure}
   \centering
   \includegraphics[width=15cm]{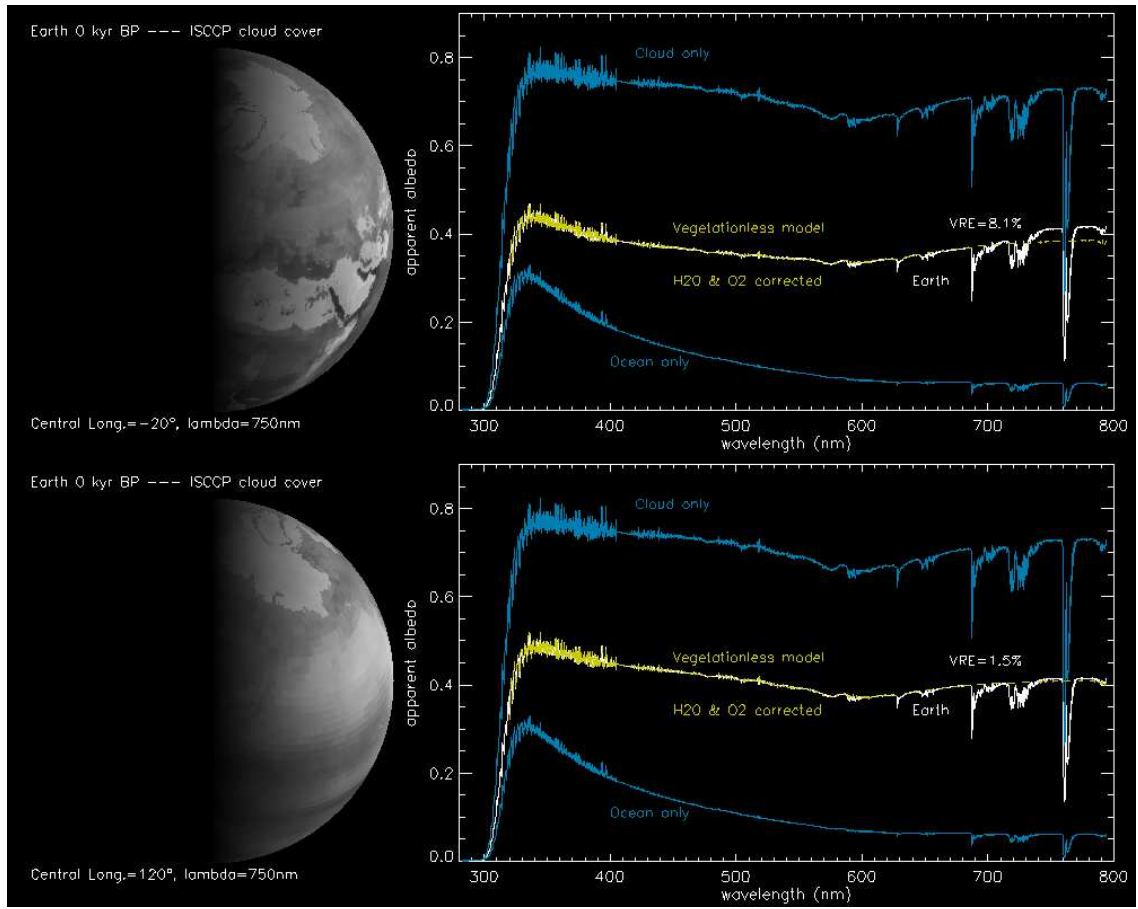}
   \caption{Earth scenes and corresponding spectra reconstructed for two observer's positions at inclination $i=45\char23$ (Table \ref{VRE-24h-variation_nord} and Fig. \ref{VRE-24h-variation_nord}), for central longitudes of $-20\char23$ (top, Africa and Europe) and $120\char23$ (bottom, Northern Pacific Ocean and Alaska) at which the model gives the maximum and minimum VRE, respectively. The Earth spectrum is in white, and a vegetationless model in yellow (moreover with $O_2$ and $H_2O$ bands corrected) allows to visualize and calculate the VRE. For the record, cloud and ocean spectra are shown in blue. Scenes are for $\lambda=750\ nm$, a wavelength at which vegetation is brighter than in the visible. For a better readability, the radiance is shown here at power 0.7 to enhance the darker regions of the scene and underline the shape of the sunlit Earth seen by the observer.}
\label{earth_scenes_spectra_45}
\end{figure}

\begin{figure}
   \centering
   \includegraphics[width=15cm]{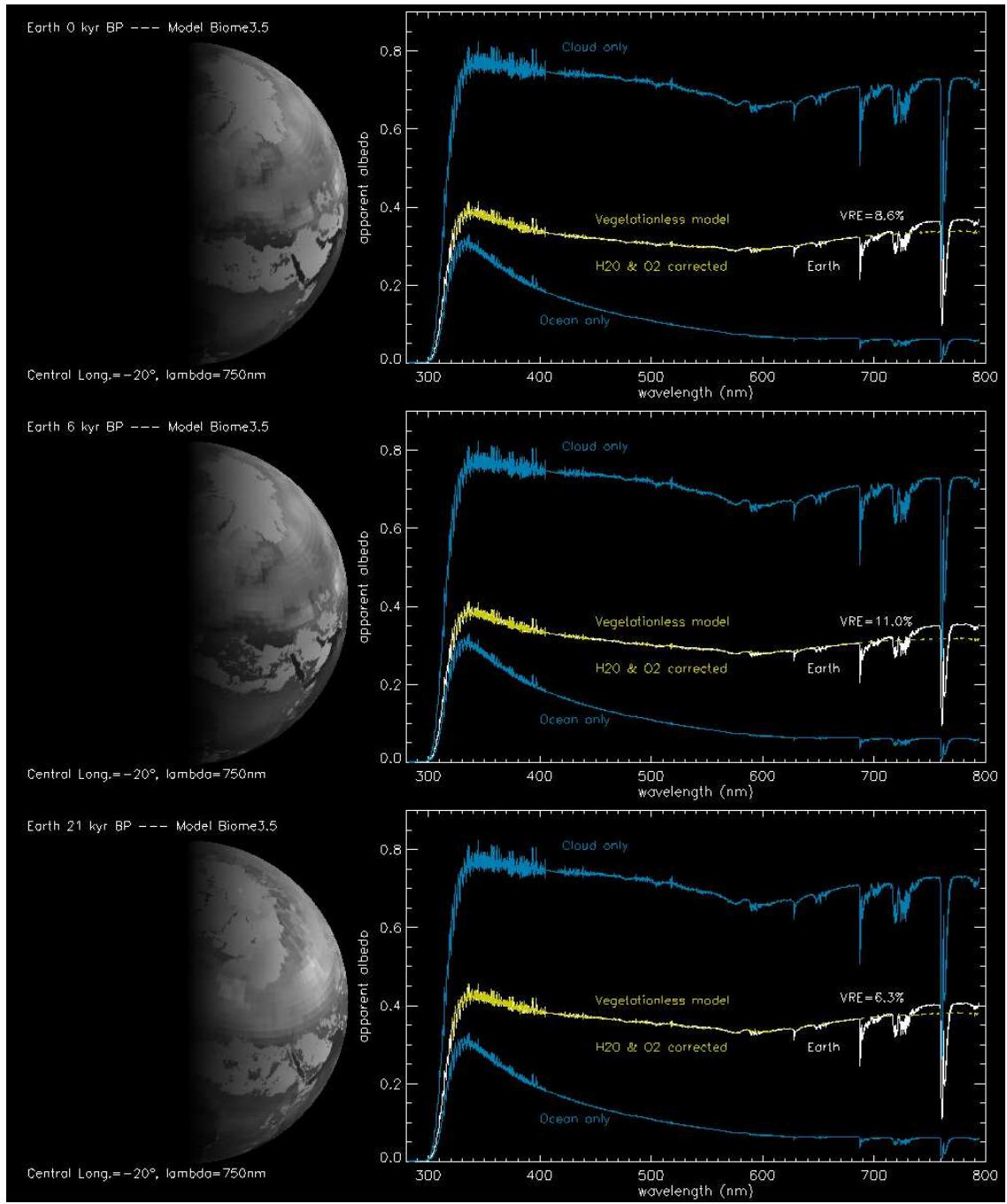}
   \caption{Earth scenes and corresponding spectra reconstructed for one observer's position at inclination $i=45\char23$ (Table \ref{VRE-24h-variation_nord} and Fig. \ref{VRE-24h-variation_nord}), for central longitudes of $-20\char23$ (Africa and Europe) at which the model gives the maximum VRE. The Earth spectrum is in white, and a vegetationless model in yellow (moreover with $O_2$ and $H_2O$ bands corrected) allows to visualize and calculate the VRE. For the record, cloud and ocean spectra are shown in blue. Scenes are for $\lambda=750\ nm$, a wavelength at which vegetation is brighter than in the visible. For a better readability, the radiance is shown here at power 0.7 to enhance the darker regions of the scene and underline the shape of the sunlit Earth seen by the observer.}
\label{earth_scenes_spectra_45_biome35}
\end{figure}

Figures \ref{earth_scenes_spectra_45} and \ref{earth_scenes_spectra_45_biome35} finally show modern Earth's scenes with corresponding spectrum obtained following the method described in Sect. \ref{method}. In Fig. \ref{earth_scenes_spectra_45}, we show the spectrum for central longitudes of $-20\char23$ (Africa and Europe) and $120\char23$ (Northern Pacific Ocean and Alaska) at which the model gives the maximum and minimum VRE, respectively of 8.1 and 1.5\%. Fig. \ref{earth_scenes_spectra_45} show the Earth spectra based on Biome3.5 data for central longitudes of $-20\char23$ (Africa and Europe) at which the model gives the maximum VRE, i.e. 8.6, 11.0 and 6.3\% at 0, 6 and 21 kyrBP, respectively.

Will next space observatories be able to detect VRE-like features in an Earth-like planet spectrum ? The Terrestrial Planet Finder Coronagraph mission (TPF-C) is aimed to make the direct detection and the characterization (multi-spectral imaging at resolution up to 70) of nearby terrestrial extrasolar planets in the visible range. The telescope would have an elliptical aperture of $(6\ \rm{to}\ 8) \times 3.5\ m^2$ \cite{stapelfeldt2005, shaklan2006}. An Earth-like planet in the habitable zone is $10^{10}$ fainter than its parent star, making such planets very faint objects in the V=28 to 31 range. According to \cite{stapelfeldt2005}, exposure time of the order of one day will be needed to detect them and spectral characterization will add 'weeks of additional integration time per target'. More quantitatively, in a comparison of different types of coronagraphs \cite{guyon2006}, the authors conclude that an Earth-like planet at about 10 pc could be detected in 2.3 h with a signal-to-noise ratio of SNR=7 with a 100 nm bandwidth from 0.5 to $0.6\ \mu m$ with a 4-m telescope, and even in 14 minutes with a 6-m telescope, both equipped with one of best performers of the comparison, a phase-induced amplitude apodisation coronagraph. Now if we want to be able to detect a 4\% VRE-like feature in a spectrum at a resolution of 70 around 700 nm, the SNR should be about 80 for a 3-sigma detection, or 125 for a 5-sigma detection. With the 4-m coronagraphic telescope metioned above, the total exposure time becomes vey long, 18 and 44 weeks for 3 and 5-sigma detection, respectively. This in agreement with the order of magnitude given by \cite{stapelfeldt2005}. With a 6-m telscope, the total exposure times decrease to 13 days and one month, respectively. If a larger telescope than the TPF-C-like telescopes mentioned above is considered, the exposure time will decrease again. According to \cite{guyon2006}, a planet could be detected in 2 minutes with a 12-m telescope. The total exposure times to detect a 4\% feature in a spectrum would then decrease to 2 and 5 days, for 3 and 5-sigma detection, respectively. These very long exposure times suggest that features of a few percents will be extremely difficult to detect, maybe even not within reach of a 'first-generation' 4-m TPF. A 6-m telescope seems to be the minimum to envisage the detection of seasonal variations. With exposure times of 2 to 5 days, a 12-m telescope would probably not freeze the rotation of the planet and rather give an integrated spectrum of the planet over several rotations, but atmospheric changes at times scales of the order of a week or longer would be possible to detect and follow up, as well as longer seasonal variations.  

Unfortunately TPF-C might not fly before 2025, and in the meantime, a smaller instrument could look for bigger and closer terrestrial planets: A 2-m class space coronagraph like SEE-COAST could indeed detect a VRE on super-Earths closer than 3 to 6 pc \cite{schneider2008}, depending on their radius whose maximum value is presently unclear. These super-Earths are as well habitable than Earth-sized planets.

\begin{figure}
\centering
\includegraphics{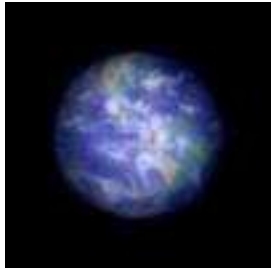}
\caption{Simulated image of the Earth at 3 pc (10 light-years) observed with a 150-km hypertelescope interferometric array made of 150 3-m mirrors working at visible wavelengths (from \cite{labeyrie99}). North and South America are visible. Note that this simulation is done at visible wavelengths, while in the (very) near-IR at 750 nm, vegetated areas would be much brighter and more easily detectable on continents. Spatial resolution at 750 nm would remain the same than at visible wavelength with the same hypertelescope flotilla being spread over 225-km instead of 150-km. }
\label{labeyrie99}
\end{figure}

More ambitious instruments that might fly later this century will allow us to make a step furthermore in the characterization of these planets and the observation of 'exo-vegetation'. Hypertelescopes in space - interferometric sparse arrays of small telescopes - will indeed allow us to see Earth-like planets as \textit{small resolved disks} several resels across \cite{labeyrie99, labeyrie2008} and this clearly will help us to detect photosynthetic life on these planets! For example, a 150-km hypertelescope would provide 40 resolution elements (resels) across an Earth at 3 pc in yellow light \cite{labeyrie99}. And a formation of 150 3-m mirrors would collect enough photons in 30-min to freeze the rotation of the planet and produce an image with at least $\approx 300$ resels, and up to thousands depending on array geometry (Fig.~\ref{labeyrie99}).  At this level of spatial resolution, it will be possible to identify clouds, oceans and continents, either barren or perhaps (hopefully) conquered by vegetation.

\begin{figure}
   \centering
   \includegraphics[width=13cm]{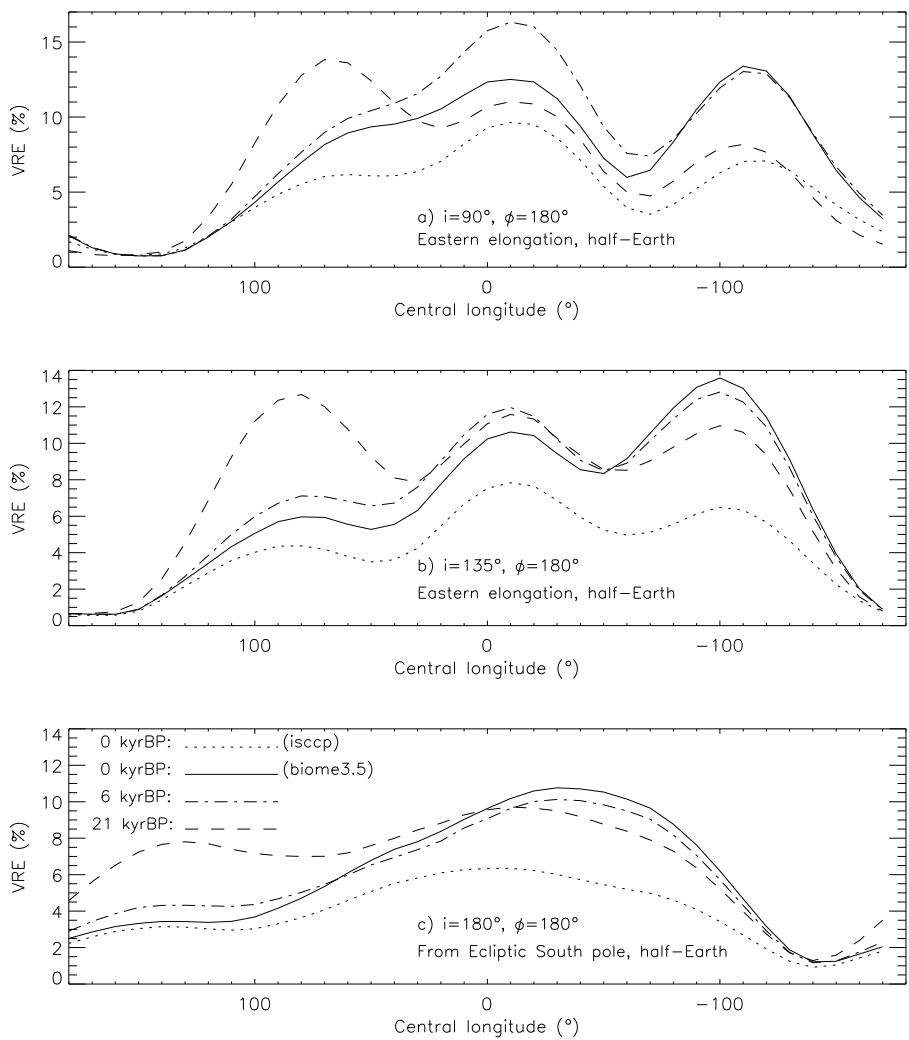}
   \caption{Vegetation Red Edge measured with Biome3.5 for clouds, ice and biomes, and GOME spectra (and ISCCP for clouds at 0 kyrBP only). The VRE calculated with the ISCCP clouds are significantly lower than those calculated from the BIOME3.5 model, probably due to the difference between the corresponding cloud maps for the southern Hemisphere as suggested by Table \ref{data_cc}.
   Central longitude is decreasing, according to the eastward Earth rotation. }
\label{VRE-24h-variation_sud}
\end{figure}

\begin{figure}
   \centering
   \includegraphics[height=18cm]{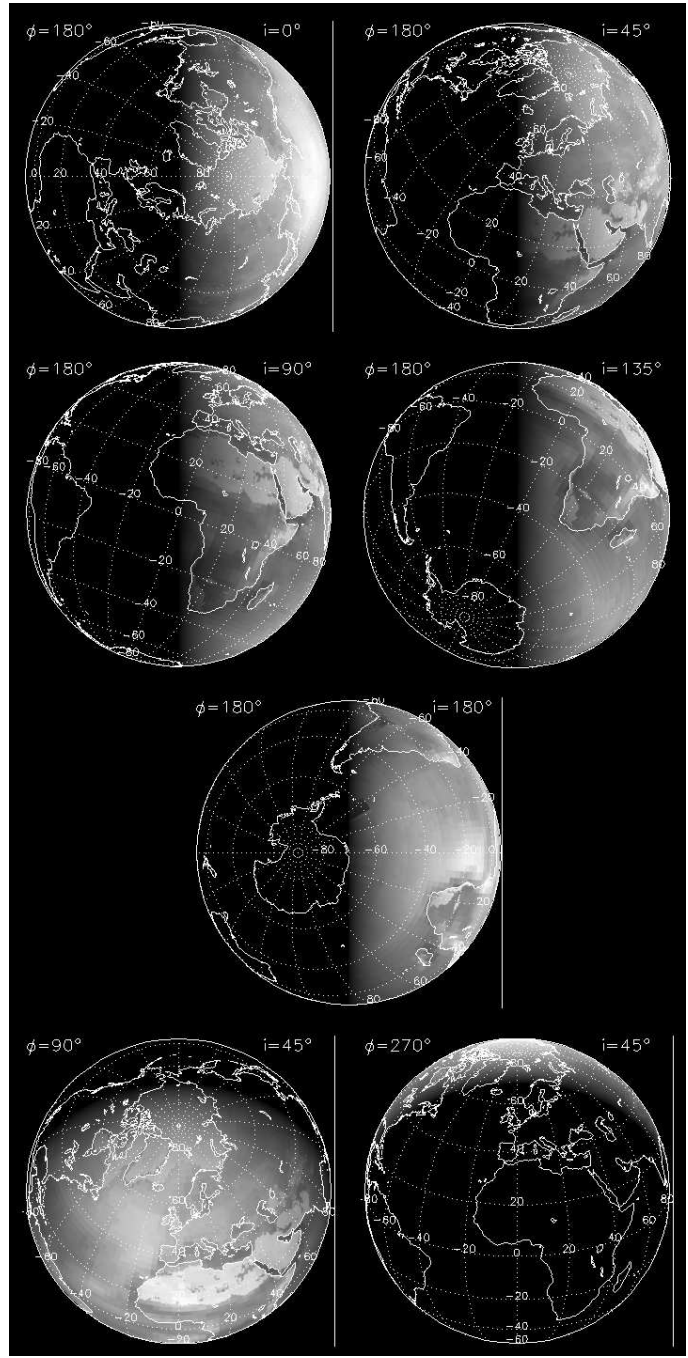}
   \caption{Earth scenes reconstructed for the observer's positions given in Table \ref{VRE_albedo}, at  $0\char23$ of central longitude. Scenes are for $\lambda=750\ nm$, a wavelength at which vegetation is brighter than in the visible. For a better readability, the radiance is shown here at power 0.7 to enhance the darker regions of the scene and underline the shape of the sunlit Earth seen by the observer.}
\label{earth_scenes_all}
\end{figure}

\section{Conclusion}
In order to mimic an Earth-like extrasolar planet observed without spatial resolution, i.e. as a single dot, in a different climate state than our contemporary Earth, we considered the Earth during the last climatic extrema, 6 and 21 kyr ago, and studied the vegetation spectral signature integrated over the full illuminated phase seen by the observer. The results presented in this paper suggest that vegetation disk-averaged signature remained visible in Earth spectra during these climatic extrema, with VRE values close to those for the modern Earth. Our discussion emphasized some differences induced for example by the greener Sahara during the Holocene optimum, or by the large northern ice cap during the LGM. Results would probably benefit in accuracy from radiative transfer calculations, an implementation of the sun glint effect and the BRDFs of all scattering surfaces included in the model. The spectra would thence also display a more accurate Rayleigh scattering 'slope' and absorption bands depth. Further, improved GCM estimates for these past periods are now available, including coupled ocean-atmosphere simulations \cite{braconnot2007}, as well as dynamic vegetation models \cite{sitch2003}, both of which may help improve the past vegetation distributions obtained here.  All these improvements could be considered for future developments. 

Of course a next step in this work could also be the consideration of the Earth at much older epochs, when continents had a very different shape than today and climates maybe much more extreme than those considered in this work, although paleoclimatic models are probably less confident for these epochs. It would although probably provide us with relevant illustrations of Earth spectrum variability over one full day, and related VRE variations. Current climate change, thence biome maps for the end of this century, might also tell us if vegetation will remain a detectable biomarker for observers out there.




\vspace{0.5cm}
\textbf{Acknowledgements}
This work was supported in part by the French CNRS \textit{GdR Exobiologie}. The authors thank Jean-Claude Lebrun (Service d'A\'eronomie, CNRS, Verri\`eres-le-Buisson, France) who
provided the MODTRAN spectra of $H_2O$ and $O_2$, and Andreas Richter (Institute of Environmental Physics, University of Bremen, Germany) for SCIATRAN
data and discussions on Rayleigh scattering.

\end{document}